\documentclass[a4paper]{article}
\usepackage{booktabs} 

\newcommand{\HDT}{{\tt Facetize}}
\newcommand{\HIPP}{{\tt Hippalus}}
\usepackage{graphicx}
\usepackage{balance}  
\usepackage[shortlabels]{enumitem}
\usepackage{multirow}
\usepackage{amssymb}
\newlist{todolist}{itemize}{2}
\setlist[todolist]{label=$\square$}

\usepackage[autostyle]{csquotes}
\setlist[enumerate, 1]{1\textsuperscript{o}}
\usepackage{url}
\usepackage{color}
\usepackage{subfig}
\usepackage{booktabs,amssymb,makecell,rotating}
\long\def\comment#1{}

\usepackage{pgfplots}
\pgfplotsset{width=7cm,compat=1.8}
\usepackage{pgfplotstable}

\usepackage{sfmath}

\usepackage{enumitem}

\usepackage{geometry}
\newgeometry{vmargin={20mm}, hmargin={18mm,20mm}}   

\begin{document}
\title{Facetize: An Interactive Tool for Cleaning and Transforming Datasets for Facilitating Exploratory Search}

\author{Anna Kokolaki \  and\  Yannis Tzitzikas \\
	Institute of Computer Science, FORTH-ICS, GREECE, and \\
	Computer Science Department, University of Crete, GREECE \\
	kokolaki@ics.forth.gr,   tzitzik@ics.forth.gr 
}

\maketitle

\begin{abstract}
There is a plethora of datasets in various formats which are usually 
stored in files, hosted in catalogs, or accessed through SPARQL endpoints.
In most cases, these datasets cannot be straightforwardly
explored by end users,
for satisfying recall-oriented information needs.
To fill this gap, 
in this paper we present
the design and implementation
of
\HDT\,
an editor that allows  
users
to transform
(in an interactive manner)
datasets,
either static (i.e. stored in files),
or dynamic
(i.e. being the results of SPARQL queries),
to datasets that can be directly explored effectively
by themselves or other  users.
The latter (exploration) is achieved   through the familiar interaction paradigm of Faceted Search 
(and Preference-enriched Faceted Search).
Specifically in this paper we describe the requirements,
we introduce the required set of transformations,
and then we
detail the functionality and the implementation of the
editor \HDT\
that realizes these transformations.
The supported operations cover a wide range of tasks
(selection, visibility, deletions, edits, definition of hierarchies, intervals, derived attributes, and others)
and \HDT\  enables  the user to carry them out 
in a user-friendly and guided manner,
without presupposing any technical background 
(regarding data representation or query languages).
Finally we present the results of an  evaluation with users.
To the best of your knowledge,
this is the first editor for this kind of tasks.
\end{abstract}


\section{Introduction}
\label{sec:INTRO}

Although there is a tendency to publish data as Linked Data, 
the majority  of datasets is still represented in simple data file formats such as CSV and TSV.
Such datasets cannot be easily exploited by end users. 
A user can view directly these CVS files using a text editor, 
or spreadsheet application,  
or on the other extreme,
one should build  dedicated applications for offering a more user friendly exploration service and experience.
To tackle this difficulty, our objective 
is to provide a general purpose solution that enables users to 
directly explore such datasets and/or to configure the way they are explored,
for better supporting recall-oriented information needs.
To this end, we rely on a quite familiar access method, specifically on {\em Faceted Search}, 
which is the defacto standard in e-commerce and booking applications.
However, a straightforward loading of such datasets in a faceted search system will not always result to a satisfying solution, 
since additional tasks are usually required. This includes deciding
(a) the parts of the dataset that should be explorable,
(b) the attributes that should be visible and their order,
(c) the transformations and/or enrichments that should be done,
(d) the groupings (hierarchical or not) of the values that should be made, (e) the addition of derived attributes, and others.
For this reason in this paper
we present the design and implementation of  an editor, called \HDT,
that allows the user to carry out these tasks
in a user-friendly and guided manner,
without the need to write any script or use any programming language.

\begin{figure}[htbp]
	\centering
	\fbox{\includegraphics[width=\linewidth]{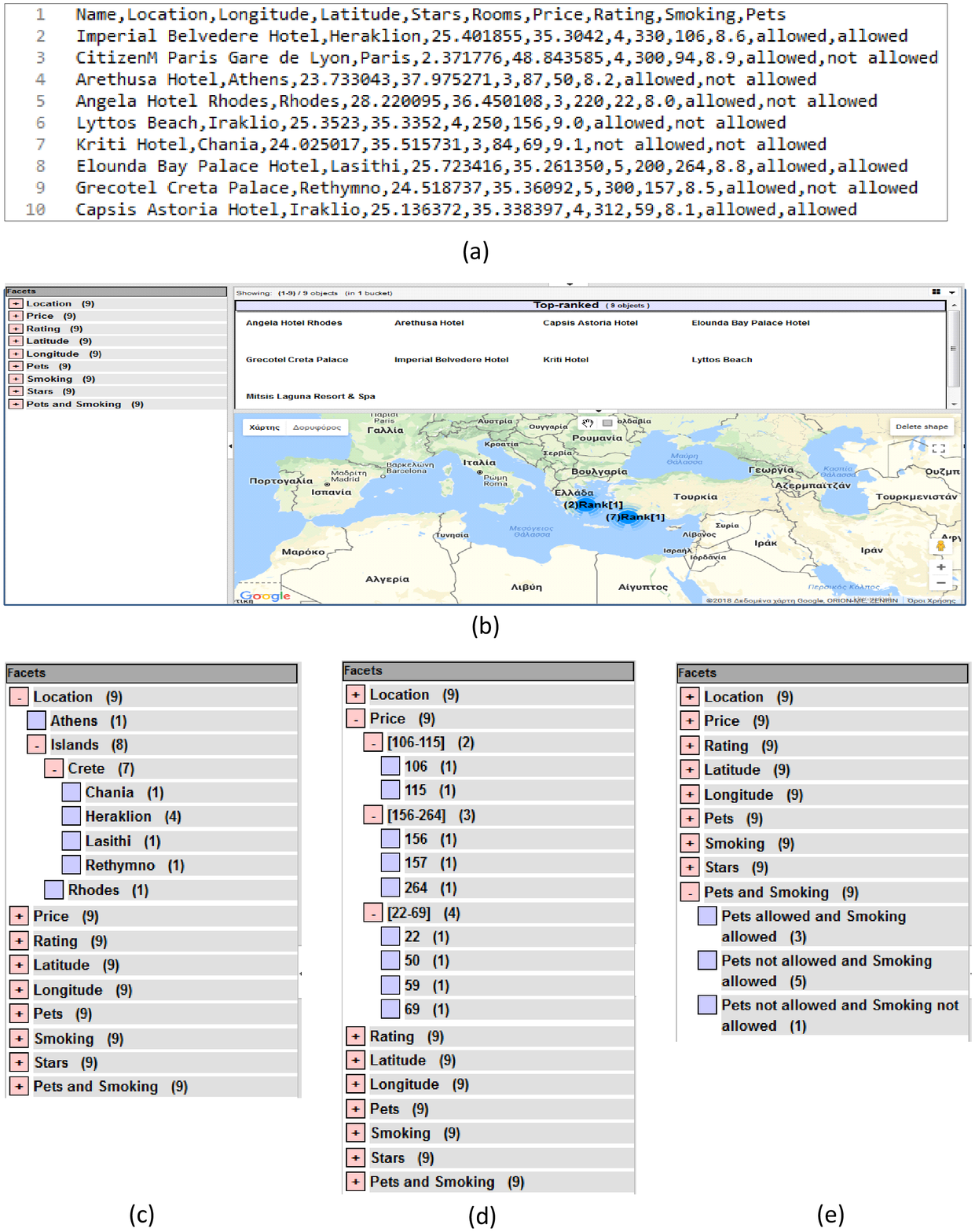}}
	\caption{Indicative Scenario}
	\label{fig:InputOutput}
\end{figure}

To grasp the idea, Figure \ref{fig:InputOutput}(a)
shows a CSV file containing information
about 9 hotels each described 
with 10 properties.
Now suppose that one would like from this dataset
to produce an explorable 
dataset,
as sketched in Figure \ref{fig:InputOutput}(b),
with the following specific requirements: \\
%
(1) The dataset must contain hotels, only located in Greece
(so one row should be deleted).\\
(2)   The dataset should be enriched with a new  hotel with values:
\textit{Mitsis Laguna Resort \& Spa,Heraklion,   25.371359,35.307237, 5,385,115,8.7, allowed, not allowed} \\
(3)  The properties with name \textit{Longitude} and \textit{Latitude}, must be marked as properties 
that contain geographic information for longitude and latitude respectively,
for enabling an exploration system 
to show the location of  each hotel on a map. \\
(4)   The hotel entities should take as names the value of the property \textit{Name}. \\
(5)   The value \textit{Iraklio} in property \textit{Location}, must be replaced with the value \textit{Heraklion}, 
in all entities that contain that value. \\
(6)  The property \textit{Rooms} should not appear in list of facets: it should either be hidden or deleted. \\
(7)  A new property with name  \textit{Pets and Smoking} should  be created,
with values as shown in the bottom part of Figure \ref{fig:InputOutput}(e),
and each hotel should be associated with the right value. \\
(8)   The values in property \textit{Location} should be organized hierarchically 
as shown in Figure \ref{fig:InputOutput}(c). \\
(9)  The values in property \textit{Price}, must be organized in interval hierarchies, 
as shown in  Figure \ref{fig:InputOutput}(d). \\
(10)  The order of the facets should be as shown in Figure \ref{fig:InputOutput}.

\ \\

\begin{figure}[htbp]
	\centering
	\includegraphics[width=120mm]{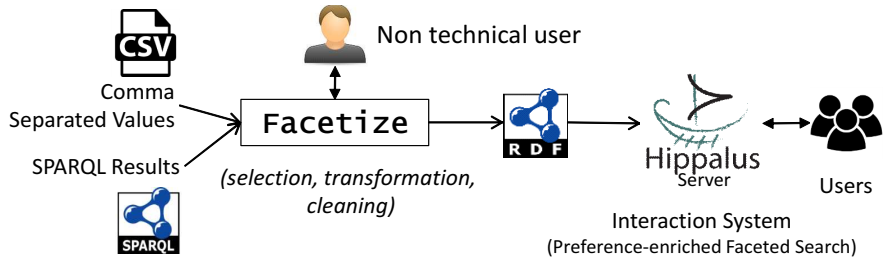}
	\caption{The process}
	\label{fig:process}
\end{figure}

With the approach that we describe in this paper,
and the tool \HDT, 
a naive user can make all these transformation in an easy manner
and produce an  output file
that is directly loadable 
to 
a system that supports Faceted Search (or Preference Enriched Faceted Search),
like \HIPP, 
which the users can directly explore 
using a GUI as shown in Figure \ref{fig:InputOutput}(b).
The overall process  is sketched in Figure \ref{fig:process}.

In a nutshell, the key contributions of this paper are:
(a) we identify the  basic requirements for this kind of tasks,
(b) we present a set of operations (for selecting, transforming, cleaning or enriching a dataset),
(c) we describe the design and implementation of  \HDT\ that  supports these operations
over static datasets (i.e. stored in files) as well as dynamic
(i.e. being the results of SPARQL queries), and
(d) we report the results of a task-based evaluation with users.
In comparison to related systems,  
OpenRefine is probably the more relevant system.
One  distinctive feature of \HDT\
is that it supports 
the creation of hierarchies and numeric intervals
and these features are crucial for managing the complexity of large datasets,
and producing datasets that can be easily explored.
Moreover, \HDT\ can fetch data directly from SPARQL endpoints,
making it appropriate for dynamic datasets.

The rest of this paper is organized as follows:
Section \ref{sec:Back}
describes background information and related work,
Section \ref{sec:Reqs} identifies the requirements,
Section \ref{sec:Editor} presents the \HDT\ editor, 
Section \ref{sec:Eval} presents the results of the evaluation,
and finally
Section \ref{sec:CR} concludes and identifies issues for further work and research.

\section{Background and Related Work}
\label{sec:Back}

\subsection{Datasets}
\label{sec:Datasets}

Although there is a tendency to publish data as Linked Data, 
e.g. see 
\cite{mountantonakis2018high},
a publishing method  that facilitates  
data linking, aggregation and integration,
and is beneficial for the preservation  of data (e.g. see \cite{CinderellaStickSpringer2018}),
a big percentage  of datasets is  represented only in simple data file formats such as CSV.
\comment{
	There has been a lot of activity in open data around the world. 
	CKAN data hubs and portals\footnote{https://ckan.org/} exist in Austria, Brazil, US, Africa and many other countries. 
	The top-five most used data categories \footnote{https://www.europeandataportal.eu/sites/default/files/re-using\_open\_data.pdf} are government and public sector (11.9\%), economy and finance (11.6\%), regions and cities (10.1\%), population and society (9.5\%), and environment (8.9\%). 
	These five data categories together represent 52\% of the total re-use of Open Data. 
	The least used data categories are international issues (3.6\%), health (4.2\%) and justice, legal system and public safety (4.2\%).
}
For instance,  in {\tt Govdata.de},
one of the most popular open data portals, 
almost 26.76\% of its files are
in CSV format\footnote{From open data formats statistics of the main catalogs (November 2016)}.
In 
{\tt Data.gov.uk}, the official open data
portal of the UK Government,
18.70\% of them are in CSV format. 
In the open data portal of Italy, {\tt Dati.gov.it},  around 29.61\% files in CSV format,
while in the {\tt european data portal.eu} (EDP)  around 11.68\% of the files are CSV files.


\subsection{Faceted Search}
\label{sec:FS}

{\em Faceted Exploration} (or {\em Faceted Search}) is a widely used interaction scheme for Exploratory Search. It is the de facto query paradigm in e-commerce \cite{sacco2009dynamic,tunkelang2009faceted}.
In a short (and  rather informal) way we could define  it as a
{\em session-based interactive method for query formulation
	(commonly over a multidimensional information space) through
	simple clicks that offers an overview of the result set  (groups and count information), never leading to empty results sets}.
Faceted search has been generalized also for  RDF datasets \cite{tzitzikas2016faceted}.


\subsection{PFS: Preference-enriched Faceted Search}
\label{sec:PFS}

\textit{Preference-enriched Faceted Search}
\cite{tzitzikas2012interactive}, for short PFS,
is an extension of Faceted Search that supports preferences.
PFS offers actions that allow the user to order facets, values, and objects using \textit{best, worst, prefer to} actions (i.e. relative preferences), \textit{around to} actions (over a specific value), or actions that order them lexicographically, or based on their values or count values. Furthermore, the user is able to \textit{compose} object related preference actions, using \textit{Priority, Pareto, Pareto Optimal} (i.e. skyline) and other.
The distinctive features of PFS
is that it allows expressing preferences over attributes whose values can be hierarchically organized (and/or multi-valued), it  supports preference inheritance,
and it offers  scope-based rules for resolving  automatically the conflicts that may arise.
As a result the user is able to restrict his current focus by using the faceted interaction scheme (hard restrictions) that lead to non-empty results, and rank according to preference the objects of his focus.
Recently, PFS has been used in various domains,
e.g.  for offering a flexible process for the identification of fish species \cite{pfs2016species},
as a Voting Advice Application \cite{voting2016aid} and it has been expanded with geographic anchors for being appropriate for the exploration of datasets that contain also geographic information \cite{lionakis2017pfsgeo}.
Applications of the model in the context of  spoken dialogue systems
are also emerging \cite{papangelis2017ld}.

In our work we decided to use
\HIPP\ 
which is  a publicly accessible web system that implements the PFS interaction model.
The information base that feeds \HIPP\ is represented in RDF/S\footnote{http://www.w3.org/TR/rdf-schema/}  
using a schema adequate for representing objects described according to dimensions with hierarchically organized values.

\subsection{Related Work}
\label{sec:RW}

The existence of anomalies in real-world data motivates the development and application of data cleansing methods. 
This is a hot topic, e.g. see \cite{abedjan2016detecting} for a recent  overview,
and there are several tools for such tasks 
including,
DataPreparator\footnote{http://www.datapreparator.com/},
Potter's Wheel \cite{raman2001potter},
WinPure \cite{housien2013comparison},
OpenRefine\footnote{openrefine.org},
Karma\cite{gupta2012karma}, 
DataX-Former \cite{abedjan2016dataxformer}, 
Katara\cite{chu2015katara},
Data Wrangler \cite{kandel2011wrangler}
and others.
Existing data cleansing approaches mostly focus on the transformation of data, the elimination of duplicates, 
syntax and lexical errors detection.  
They also offer methods for create and edit attributes using a number of functions.  
In general, the users can see the transformations in each step of editing and 
they can export the data after applying the transformations 
in various file formats, such as Excel, CSV, RDF and others.

We could say that the the available data cleaning solutions and tools 
(\cite{abedjan2016detecting},  \cite{maletic2005data})
belong to one or more of the following four categories:
\begin{itemize}
	\item	\textit{Rule-based detection algorithms} that can be embedded into frameworks and the user can specify a collection of rules that clean data will obey and the tool will find any violations.
	\item	\textit{Pattern enforcement and transformation tools} such as OpenRefine\footnote{http://openrefine.org/}, Data Wrangler \cite{kandel2011wrangler}, Trifacta\footnote{http://www.trifacta.com}, Katara\cite{chu2015katara}, and DataX-Former \cite{abedjan2016dataxformer}. These tools discover patterns in the data, either syntactic (e.g., OpenRefine and Trifacta) or semantic (e.g., Katara), and use these to detect errors (cells that do not conform with the patterns). They can also be used to change data representation and expose additional patterns.
	\item		\textit{Quantitative error detection algorithms} that expose outliers, and glitches in the data.
	\item		\textit{Record linkage and de-duplication algorithms} for detecting duplicate data records, such as the Data Tamer system \cite{stonebraker2013data}. These tools perform entity consolidation when multiple records have data for the same entity. Conflicting values for the same attribute can be found, indicating possible errors.
\end{itemize}


\section{Requirements}
\label{sec:Reqs}

We dichotomize requirements 
to transformation requirements 
and
to workbench-related requirements.
They are presented in a more structured way, later in Table \ref{tab:table1}
of Section \ref{sec:RT}.

\subsection{Transformation Requirements}
\label{sec:TR}

Even from the running example in the introductory section 
it is evident that for preparing a dataset appropriate for exploration
one  should be able to define  
the facets that should be visible,
and their order.
The user  should be able to  specify the type of the terms  of a facet,
e.g. identifier, integer, float, string,  longitude and latitude.
By defining the type of a facet
as identifier, 
the entities will take as names their corresponding values in this facet, 
and  all  values in this facet should be  distinct.
Moreover it should be possible  to define  new facets
whose terms are derived  by applying functions over the terms of other facets.
For avoiding cluttering the GUI,
and aiding the exploration,
it  should be possible
to define hierarchical groupings  of terms,
as well as intervals to numerical values.
Finally,
it should be able to  
delete all rows with a specific value or a condition,
to replace each distinct value in a facet with a new one in each row of the dataset,
and 
the user should also be able
to 
add, edit and delete  individual rows.

\comment{==
	In this section we will describe the transformations supported by \HDT\ that users could apply to datasets. 
	Specifically, user would be able to specify the  \textbf{order} and  \textbf{visibility} of facets. Also, there is the ability to organize values of facets in \textbf{hierarchies}. There are supported operations for move values in existing hierarchies, add parent values in other hierarchies or facets’ values. In case of string values, user could use functions to define expressions for create groups of values with same prefix and values that start with the same range of letters  for aiding the exploration (and for avoiding cluttering the GUI). In the case that there are number values in a facet, user could also define linear, logarithmic \textbf{intervals} and intervals with specific bounds.
	Moreover, user would be able to define \textbf{derived} facets, using a range of functions to create expressions for facets' values.
	
	Also, user could specify the \textbf{type} for each facet: integer, float, string, identifier, boolean and specify facets as containing information for geographic data: latitude, longitude. 
}

\subsection{Workbench-related Requirements}
\label{sec:WR}

The notion of project, should be supported, 
allowing the user to create,  open, edit and save the changes on disc.
Moreover the system should be able to keep the history of  transformations that have been applied
and offer to the user the ability to undo  the desired ones (and redo).
Due to data dynamicity, 
the user  should be able to open a project 
and change  the input dataset, i.e. by giving a more recent version of the dataset, 
or the file with the hierarchical information.
This is very important for datasets that change over time,
since  we would not like to loose the transformations that have been 
defined for a past version of the dataset.
Finally, it should be possible to  export the transformed dataset
in RDF according to a schema that is compliant with \HIPP,
for enabling the straightforward loading by \HIPP.

\section{The Tool \HDT}
\label{sec:Editor}

\comment{
	Here we discuss the implementation of \HDT\ (in \S \ref{sec:Implementation}),
	the supported transformations (in \S \ref{sec:SupportedTranformations}),
	the notion of project (in \S \ref{sec:Project}), 
	an indicative interaction example (in \S \ref{sec:INTER}), and 
	then (in \S \ref{sec:ERRORS})
	how it tackles the error correction requirements.
}

\begin{figure}
	\centering
	\subfloat[]{%
		\begin{minipage}{\linewidth}
			\includegraphics[width=0.50\linewidth]{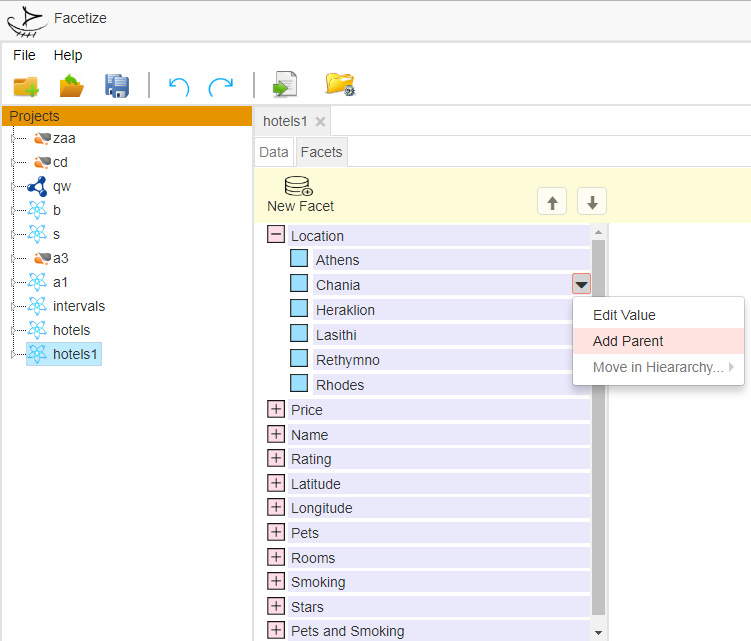}\hfill
			
		\end{minipage}%
	}\par
	\subfloat[]{%
		\begin{minipage}{\linewidth}
			\includegraphics[width=0.40\linewidth]{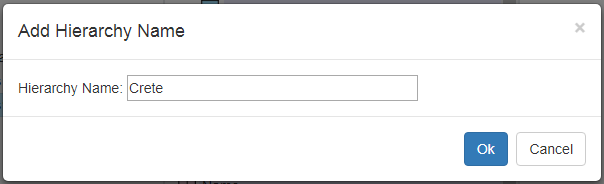}\hfill
			\
		\end{minipage}%
	}\par
	\subfloat[]{%
		\begin{minipage}{\linewidth}
			\includegraphics[width=0.50\linewidth]{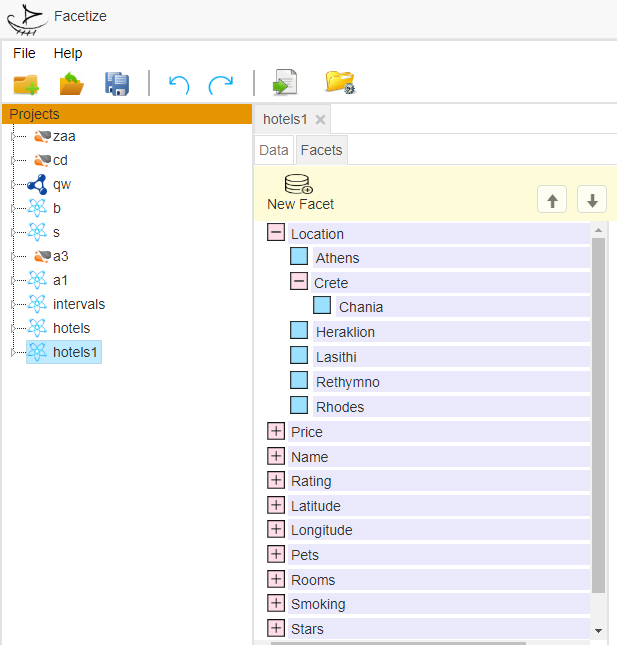}\hfill
			
		\end{minipage}%
	}
	\caption{(a) Click on a term and open menu, (b) Set parent $Crete$ for term $Chania$ in the hierarchy, (c) Hierarchy Chania/Crete has been created}
	\label{fig:interactionExample}
\end{figure}

\noindent {\bf Supported Transformations.}
In brief, \HDT\  supports all transformation requirements
described in 
\S \ref{sec:Reqs}.
As regard hierarchies, 
the user is able
to  move values in existing hierarchies, 
and add intermediate terms.
In case of string values, 
the user  is able 
to use functions to define expressions for creating groups of values 
with same prefix and values that start with the same range of letters  for aiding the exploration.
For numerical facets, the specification of intervals is supported
and the user can  define linear intervals, logarithmic intervals, and intervals with specific bounds.
%
%
For example, 
Figure \ref{fig:interactionExample}.a shows the list of facets 
of the running example described in the introductory section.
Notice that facet $Location$ contains 6 distinct values. 
By left clicking on the term $Chania$, 
and selecting the  option $Add\ Parent$ from the menu, 
a (bootstrap.js) modal dialogue appears  with a text field
for adding  the  parent value (see Fig. \ref{fig:interactionExample}.b). 
If for example the user inserts the  value $Crete$, 
the hierarchy Chania/Crete will be created in facet $Location$ (as shown in Fig. \ref{fig:interactionExample}.c).

In terms of the data error types,
\HDT\ assists the user to  detect and remove them.
Specifically, 
in case of \textit{Rule}, \textit{Pattern violations} and \textit{Outliers} errors in the dataset, 
the user can edit the values that cause the inconsistencies and replace them with other values. 
In case of \textit{Duplicates}, the user can delete or edit duplicate rows. 

\ \\
\noindent
{\bf The Notion of Project.}
\HDT\ supports the notion of project, 
allowing the user to gradually configure the desired presentation
and also to update (refresh) the underlying dataset
without losing the transformations that  have been defined.
Each project has a name, an input dataset and a configuration 
of the transformations that should be applied for producing an output 
dataset that is suitable for exploration by \HIPP.
In brief, \HDT\ supports three types of project: 
(a) Project with Single File Dataset, 
(b) Project with Multiple Files, 
(c) Project with SPARQL Query.

\ \\
\noindent {\bf (a) Project with Single File Dataset}
The first row in a dataset file is the header row. Each column
in this row must contain a Property name (the name of the corresponding facet).
Hierarchies can be specified 
(a) internally, 
(b) externally, 
or 
(c) by both methods. 
According to method (a), 
the user can specify the complete path from the leaf term to its root, e.g. Mazda/Japanese/Asian. 
According to method (b), the hierarchy is specified through a configuration file that contains hierarchical information about
some or all the values of the tabular data file,
i.e. each line contains hierarchical information about one 
term using the following syntax:
name + path from the specific term to the root of the hierarchy separated with slashes (/), 
e.g. Mazda/Japanese/Asian/Manufacturer, where Manufacturer is the name of the facet that the mentioned hierarchy belongs. 
The order of rows is not important and the import of a configuration file in each project is optional, 
since  the user can create the desired hierarchies though the editor.

\ \\
\noindent
{\bf  (b) Project with Multiple Files.}
\HDT\  supports input from multiple CSV or TSV files, located at a specified folder. 
This structure is convenient for enriching the dataset with more dimensions.
The files that such project can contain are:\\
{\em Object Id File}: 	
The object id file is specified explicitly by the user and is the one that holds all the  ids 
of the objects of this project. 
It is a CSV/TSV file, with a single column,
containing the  names of the entities. 

\noindent
{\em Dimension (Property) Files}.
These files have 2 columns: the first is for the object id 
while the second for the value that the specified entity has for this property.
The hierarchies can be expressed anywhere  in the file (at the beginning, end, or mixed)
and they  can be straightforwardly represented:
we just have to put the child at the first column and its parent at the second column. 
The translator detect that a property is hierarchical when the name in the first column is not a valid object id.

\ \\
\noindent 
{\bf (c) Project with SPARQL Query.}
For  exploiting the wealth of Linked Data and the available SPARQL endpoints,
\HDT\ allows
the user to specify the address of a SPARQL endpoint and 
the  SPARQL SELECT query to be sent,
e.g. the query
{\tt SELECT ?Speed ?Price ?Weight}
will lead the tool  to create 3 properties: Speed, Price and Weight.
Moreover the user can save queries and endpoints in the favourites list,
for re-running them in the future.
\comment{
	Also note that if the SPARQL endpoint 
	supports SPARQL-LD \cite{fafalios2015sparql,fafalios2016querying},
	then  that implies
	that \HDT\ projects can be defined
	by extracting information from  RDF dumps and  HTML pages with RDFa. 
}

\ \\
\noindent {\bf Project Refreshing.}
Every time the user applies a transformation to the dataset of a \HDT\ project, 
that transformation is saved in JSON format in a file  in the project's folder. 
For each transformation,  its type, as well as all related information, are saved,
for enabling reapplying the transformation in the future.

\ \\
\noindent {\bf Implementation.}
%
\HDT\  is a Web Application, based on the architecture of client-server.
The presentation layer ($front\ end$)
is  implemented using HTML, CSS for rendering elements appropriately on screen,
and JavaScript libraries such as 
{\tt Bootstrap.js} to create responsive content. 
The data access layer ($back\ end$), 
is implemented using Java Servlets technologies 
and
Jena Semantic Web framework\footnote{http://jena.apache.org/}. 
To run the editor
the user should  
have installed at his computer 
Java Runtime Environment (JRE)
and  a Java Servlet Container.


\comment{ = To parakatw den xreiazetai se paper ==
	It was implemented using NetBeans IDE\footnote{https://netbeans.org/} development environment. User can run locally, from command line the .war file that is being exported from NetBeans, using a java Web Server.
	=}

\renewcommand*\theadfont{\bfseries}
\settowidth\rotheadsize{\theadfont Infrastructure}
\renewcommand\theadgape{}
\renewcommand\theadalign{lc}
\renewcommand\rotheadgape{}
\begin{table*}[htbp]
	{\small
		\centering
		\caption{Features of data cleansing systems}
		\label{tab:table1}
		\begin{tabular}{|p{3.4cm}|p{8cm}||c|c|c|c|c||c|}
			\toprule
			\thead{Category} &{Feature} & \rothead{DataPreparator} & \rothead{Potters Wheel} & \rothead{WinPure Clean}& \rothead{Karma} & \rothead{OpenRefine}  & \rothead{Facetize}  \\\hline\hline

			\multirow{4}{*}{Import/Export}
			&Import/Export text files, EXCEL & $\checkmark$ &   & $\checkmark$ & &$\checkmark$ &$\checkmark$ \\
			&Import/Export Data from databases   & $\checkmark$ &  & $\checkmark$ & & & \\
			&Import/Export RDF files & &   & & $\checkmark$&$\checkmark$&$\checkmark$\\
			&Export R2RML/JSON files & &   & &$\checkmark$ & &\\
			&Execute SPARQL queries to retreive data from a source     & &   &  & & &$\checkmark$ \\\hline

			
			\multirow{4}{*}{Data cleaning: Rows}
			&Deletion of rows   & &   &  & & $\checkmark$  &$\checkmark$\\
			&Removal of duplicate rows  & &   & $\checkmark$& & $\checkmark$ &\\
			&Addition of rows  & &   & &$\checkmark$ & $\checkmark$ & $\checkmark$\\	
			&Removal of empty rows  & &   & &  &$\checkmark$ &$\checkmark$\\\hline
			&Deletion of records containing  missing values  &$\checkmark$ &   &  & &  &$\checkmark$\\
			
			\multirow{6}{*}{Data cleaning: Values}
			&Character removal, text replacement, date conversion  &$\checkmark$ &   &  & & & \\
			&Value editing    & &   &  &$\checkmark$ & $\checkmark$  &$\checkmark$\\
			&Impute missing values  &$\checkmark$  &   & $\checkmark$& & $\checkmark$ & \\
			&Creation of value  hierarchies      & &   &  & &  &$\checkmark$ \\
			&Creation of intervals for arithmetic values     & &   &  & &  &$\checkmark$ \\
			&Correction of bad values using string similarity metrics     & &   &  &$\checkmark$ &  & \\
			&Clustering of the values in a property that contain specific characters (e.g.	same prefix) 
			& &   &  & &$\checkmark$   &$\checkmark$  \\\hline
			
			\multirow{10}{*}{Data cleaning: Columns}
			&Delete/move attributes  & $\checkmark$&   &  & &  &$\checkmark$  \\ 
			&Filtering   & &   & $\checkmark$ & & $\checkmark$&$\checkmark$ \\
			&Renaming of  Columns  & &   &  &$\checkmark$ & $\checkmark$&$\checkmark$ \\
			&Set column types  & &   &  &$\checkmark$ &$\checkmark$&$\checkmark$ \\
			&Creation of new columns using expression  & &  &  & &$\checkmark$  &$\checkmark$ \\
			&Split of columns   & &  $\checkmark$  &  &$\checkmark$ & &\\
			&Deletion of columns   & &  $\checkmark$  &  & & $\checkmark$  &$\checkmark$\\
			&Rearrangement of columns    & &   &  & & $\checkmark$  &$\checkmark$\\
			&Hiding of  columns    & &   &  & & $\checkmark$  &$\checkmark$\\
			&Addition/merging of columns     & &  $\checkmark$  &  &$\checkmark$ &  &\\
			&Split records into columns   & $\checkmark$&  $\checkmark$ &  & & & \\\hline

			\multirow{3}{*}{Editor Facilities}
			&Undo/redo transformations  &  & $\checkmark$  & & &$\checkmark$&$\checkmark$ \\		
			&Display the distinct property values and their frequency
			& &   & & &$\checkmark$ &\\
			&Ability to reapply same range of transformations    & &   &  & & &$\checkmark$ \\\hline
			
			\multirow{2}{*}{Other Export Functions}	
			&Export macro or a C program, or a Perl program  & &   $\checkmark$ & & & &\\
			&Extraction of Entities  & &   & & $\checkmark$ & &\\
			
			\bottomrule
		\end{tabular}
	}	
\end{table*}

\section{Evaluation}
\label{sec:Eval}

\comment{
	Section
	\ref{sec:ComparisonWithOtherSystems}
	compares the functionality of \HDT\ with the functionality of related systems.
	Section \ref{sec:Efficiency} discusses the scalability and efficiency of \HDT.
	Finally, Section \ref{sec:TEU} reports the results of an evaluation with users.
}

\subsection{Comparison with Related Tools}   
\label{sec:RT}

\comment{  
	In comparison to the above,
	\HDT\ also allows the user to create
	hierarchies of values and numeric intervals. 
	The user is also able to execute
	SPARQL queries to retrieve the data from a source, delete rows that have specific values using functions,
	and reapply the same range of transformations. 
}

Table \ref{tab:table1}, shows the features of each data cleansing system 
that was mentioned  in \S \ref{sec:RW}, as well as of \HDT.
We observe that one  distinctive feature of \HDT\
is that it supports 
the creation of hierarchies and numeric intervals
and these features are crucial for managing the complexity of large datasets,
and producing datasets that can be easily explored.
Moreover, \HDT\ can fetch data directly from SPARQL endpoints,
making it appropriate for dynamic datasets.
Finally, users have the ability to reapply the same range of transformations to an existing project and 
this is very convenient in case they want to refresh the dataset of a project 
(with a new version of the dataset)
and would like to apply the same transformations.

We tried to carry out scenario of the introductory section,
using the tools mentioned in Table \ref{tab:table1},
i.e. 
DataPreparator,
Potters Wheel,
WinPure Clean,
Karma,
OpenRefine, Facetize.
We wanted to check what transformations were feasible to perform
and how much time it took to make them.
Of course none of these tools (apart from \HDT) can produce an RDF file
that is directly loaded to Hippalus,
however we wanted to test the rest aspects of the scenario. 
With DataPreparator we managed  to perform requirements 5 and 6 (as numbered in the introductory section)
in 1 (one) minute. 
With Potters Wheel we managed to perform requirements 6, 7 in 2 (two) minutes. 
Using WinPure Clean we did not manage to satisfy any of the requirements. 
With Karma we managed to perform requirements 2, 5, 7 in 3 (three) minutes. 
Finally with OpenRefine we managed to perform requirements 1, 2, 5, 6, 7, 10 in 6 (six) minutes,
i.e. it is the tool that covered most of the requirements.

\subsection{Task-based Evaluation with Users}
\label{sec:TEU}

We conducted a task-based evaluation with users  
for getting general and specific feedback,
and for  testing the usability of \HDT\ and user satisfaction.
We used the scenario described in Section \ref{sec:INTRO} with the dataset of hotels of Figure \ref{fig:InputOutput}(a).
We prepared a simple text tutorial of 45 slides, 
a questionnaire and a file with the description of the aforementioned scenario. 
We invited by email various persons to participate in the evaluation voluntarily. 
The users were asked to carry out  the tasks and to fill the questionnaire. 
It was stated to them clearly, that they should not rush up. 
The participation to this evaluation was optional. 
Twenty persons (20), eventually participated. 
The number was sufficient for our purposes, since according to \cite{faulkner2003beyond} 20 evaluators are enough for getting more than 95\% of the usability problems of a user interface. 
In numbers, the participants were 11 (55\%) female and 9 (45\%) male, with ages ranging from 18 to 64 years. 
As regards occupation and skills, users have studied Computer Science.
In detail, 
5 (25\%) were undergraduate students, 
12 (60\%) of them postgraduate students and 
3 (15\%) computer engineers and researchers. \\

We used a questionnaire that users had to fill it in, and then send it back to us  by email. 
The questionnaire is shown below, enriched with 
the results of the survey  in the form of percentages written in bold.

\vspace*{-3mm}
\begin{itemize}
	\item {\em How many mistakes  did you make ?}
	No mistake (\textbf{20\%}),
	1-3 mistakes (\textbf{45\%}),
	3-6 mistakes (\textbf{15\%}),
	6-9 mistakes  (\textbf{5\%}),
	More than 9 mistakes  (\textbf{15\%})
	\item {\em How much time you spent for carrying out the task?}
	6-7 minutes  (\textbf{0\%}),
	7-10 minutes (\textbf{20\%}),
	10-15 minutes (\textbf{25\%}),
	15-20 minutes (\textbf{45\%}),
	More than 20 minutes (\textbf{10\%})
	
	\item	{\em Have you ever used data transforming systems like Facetize?}
	Yes, I have used systems like Facetize. (\textbf{25\%}),
	No, I have not used any system like Facetize. (\textbf{75\%})
	\item	{\em What was the final outcome of the scenario?}
	
	I successfully completed the entire scenario, and the data are displayed exactly as 
	it was requested. (\textbf{70\%}),
	I did the whole script, except for some requirements that I did not perform successfully and did not receive all the correct results from Hippalus. List the numbers (in the order in which they appear) of the requirements you did not successfully execute. (\textbf{20\%}),
	I quitted the task in the requirement with number...... (fill in the number of requirement you were trying to perform or whatever you did). (\textbf{10\%})
	
	\item	{\em How would you rate \HDT\ as a data transformation system?}
	Very Useful (\textbf{40\%}),
	Useful (\textbf{60\%}),
	Little Useful (\textbf{0\%}),
	Not Useful (\textbf{0\%})
	
\end{itemize}

\begin{figure}

	\begin{tikzpicture}
	\centering
	\begin{axis}[
	ybar, axis on top,
	height=4cm, width=8.5cm,
	bar width=0.4cm,
	ymajorgrids, tick align=inside,
	major grid style={draw=white},
	enlarge y limits={value=.1,upper},
	ymin=0, ymax=40,
	axis x line*=bottom,
	axis y line*=right,
	y axis line style={opacity=0},
	tickwidth=0pt,
	enlarge x limits=true,
	legend style={
		at={(0.5,-0.2)},
		anchor=north,
		legend columns=-1,
		/tikz/every even column/.append style={column sep=0.5cm}
	},
	ylabel={User Percentage (\%)},
	symbolic x coords={
		7-10,10-15, 15-20,
		$>$ 20},
	xtick=data,
	nodes near coords={
		\pgfmathprintnumber[precision=0]{\pgfplotspointmeta}
	}
	]
	\addplot [draw=none, fill=blue!30] coordinates {
		(7-10,20)
		(10-15, 30) 
		(15-20,20)
		($>$ 20,0) 
	};
	\addplot [draw=none,fill=red!30] coordinates {
		(7-10,0)
		(10-15, 10) 
		(15-20,5)
		($>$ 20,5) 
	};
	\addplot [draw=none, fill=green!30] coordinates {
		(7-10,0)
		(10-15, 5) 
		(15-20,0)
		($>$ 20,5) 
	};
	
	\legend{correct,includes errors,quit}
	\end{axis}
	\end{tikzpicture}

	\begin{tikzpicture}
	\centering
	\begin{axis}[
	ybar, axis on top,
	height=4cm, width=8.5cm,
	bar width=0.4cm,
	ymajorgrids, tick align=inside,
	major grid style={draw=white},
	enlarge y limits={value=.1,upper},
	ymin=0, ymax=40,
	axis x line*=bottom,
	axis y line*=right,
	y axis line style={opacity=0},
	tickwidth=0pt,
	enlarge x limits=true,
	legend style={
		at={(0.5,-0.2)},
		anchor=north,
		legend columns=-1,
		/tikz/every even column/.append style={column sep=0.5cm}
	},
	ylabel={User Percentage (\%)},
	symbolic x coords={
		no error, 1-3, 3-6, 6-9,
		$>$ 9},
	xtick=data,
	nodes near coords={
		\pgfmathprintnumber[precision=0]{\pgfplotspointmeta}
	}
	]
	\addplot [draw=none, fill=blue!30] coordinates {
		(no error,20)
		(1-3,35) 
		(3-6,10)
		(6-9,0)
		($>$ 9,5) 
	};
	\addplot [draw=none,fill=red!30] coordinates {
		(no error,0)
		(1-3,10) 
		(3-6,5)
		(6-9,5)
		($>$ 9,0)  
	};
	\addplot [draw=none, fill=green!30] coordinates {
		(no error,0)
		(1-3,0) 
		(3-6,0)
		(6-9,0)
		($>$ 9,10)  
	};
	
	\legend{correct,includes errors,quit}
	\end{axis}
	\end{tikzpicture}
	\caption{Top: Dedicated time-Success percentage. Bottom: Errors-Success percentage}
	\label{sec:ESP}
	\vspace*{-3mm}
\end{figure}
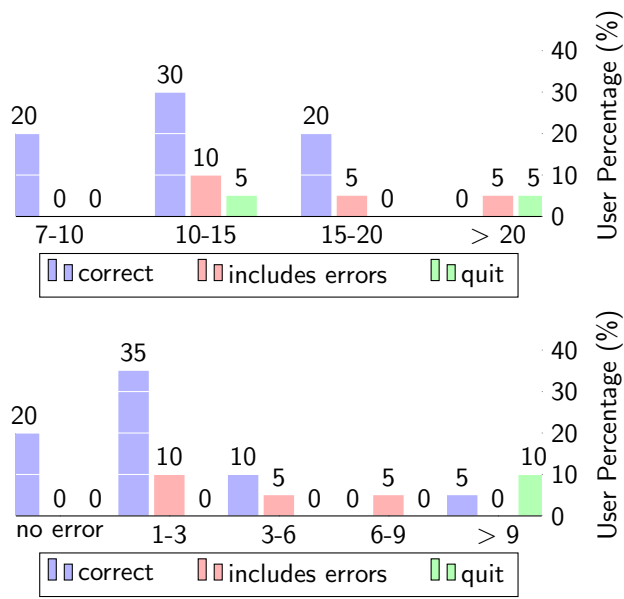

The results are quite satisfactory, 
since 14 (70\%) users managed to complete the task successfully, 
4 (20\%) executed the scenario, except from some requirements that they 
probably did not perform successfully, 
and only 2 (10\%) quitted the task. 
For those users who did not get the right results, 
or quitted the process, they did not understand the system and used it wrongly.
As regards the overall rating, 
12 (60\%) users rated the system \textit{Useful}, 
while (40\%) \textit{Very Useful}. 
Here we include only two plots from the analysis of the results.
In Figure \ref{sec:ESP}
we can see the 
``Dedicated time-Success percentage"
where we  observe that that users that spent 10-15 minutes were the most successful.
\comment{
	We observe thatn that users that spent 10-15 minutes were the most successful. 
	Those who spent more than 20 minutes were not very successful; probably they followed a wrong path or they had difficulties with the system. In contrast, those who spend 7-10 and 15-20 minutes had 20\% success. 
}
From the "Errors-Success percentage"
we can see that the users who made 1-3 errors were the most successful; 35\% success,
those who made no error had 20\% success, 
those with 3-6 errors had 10\% success, and finally
those who made more than 9 errors had 5\% success.


\section{Concluding Remarks}
\label{sec:CR}

To conclude, there is a demand for systems that help users 
with no particular technical background
to clean and apply transformations on 
plain datasets
for turning them easily explorable.
However, a straightforward loading of such datasets in a faceted search system 
will not always result to a satisfying solution, 
since additional tasks are usually required. 
This includes deciding
(a) the parts of the dataset that should be explorable,
(b) the attributes that should visible and their order,
(c) the transformations and/or enrichments that should be done,
(d) the groupings (hierarchical or not) of the values that should be made, 
(e) the addition of derived attributes, and others.
We presented the design and implementation of  an editor, called \HDT,
that allows the user to carry out these tasks
in a user-friendly and guided manner, without presupposing any technical background 
about the data representation language or the query language.
\HDT\ can be construed as an  authoring environment for setting up 
exploration services  over  static files (represented in various formats) or  results produced by querying 
SPARQL endpoints. It is worth noting that the authoring environment, supports the notion
of project,  allowing the user to gradually configure the desired presentation
and also to update (refresh) the underlying dataset
without losing the transformations that they have been defined.
We evaluated the \HDT\ prototype with users over real data
and the results of the evaluation  were  positive. 
Overall the combination of \HDT\  with   Hippalus offer a handy approach for supporting exploratory search over static and dynamic datasets. \\
We plan to make \HDT\ publicly available soon.
Directions  that are worth further work and research include 
methods for predicting the missing values.
Such functionality is helpful in cases where  there are a lot of missing values 
that we would like to fill in the dataset at hand,
and various models could be employed for that
purpose (e.g. dependence tree, Naive Bayes model, etc). 
Another direction is to device methods that can spot possible errors or misspellings
by exploiting string similarity-based clustering methods.

\comment{
	Another issue is the problem with data that contains bad values and occur for example, when people that are responsible with inserting data might be in a hurry. For correcting this type of possible errors, it\textsc{\char13}s necessary to implement a feature that indicates which string is close related to a reference one. So, using string similarity algorithms to generate the suggestions for these values, would help users to prepare their data for \textit{faceted exploration} and further usage. Also, using these string metrics, users can create group of values that have the highest similarity. Moreover, it would be useful a feature that would extract data, persons and places from one of the columns.
}

\balance

\small

\bibliographystyle{plain}
\bibliography{Facetize2018_coRR}

\begin{thebibliography}{10}

\bibitem{abedjan2016detecting}
Ziawasch Abedjan, Xu~Chu, Dong Deng, Raul~Castro Fernandez, Ihab~F Ilyas,
  Mourad Ouzzani, Paolo Papotti, Michael Stonebraker, and Nan Tang.
\newblock Detecting data errors: Where are we and what needs to be done?
\newblock {\em Proceedings of the VLDB Endowment}, 9(12):993--1004, 2016.

\bibitem{abedjan2016dataxformer}
Ziawasch Abedjan, John Morcos, Ihab~F Ilyas, Mourad Ouzzani, Paolo Papotti, and
  Michael Stonebraker.
\newblock Dataxformer: A robust transformation discovery system.
\newblock In {\em Data Engineering (ICDE), 2016 IEEE 32nd International
  Conference on}, pages 1134--1145. IEEE, 2016.

\bibitem{chu2015katara}
Xu~Chu, John Morcos, Ihab~F Ilyas, Mourad Ouzzani, Paolo Papotti, Nan Tang, and
  Yin Ye.
\newblock Katara: A data cleaning system powered by knowledge bases and
  crowdsourcing.
\newblock In {\em Proceedings of the 2015 ACM SIGMOD International Conference
  on Management of Data}, pages 1247--1261. ACM, 2015.

\bibitem{faulkner2003beyond}
Laura Faulkner.
\newblock Beyond the five-user assumption: Benefits of increased sample sizes
  in usability testing.
\newblock {\em Behavior Research Methods, Instruments, \& Computers},
  35(3):379--383, 2003.

\bibitem{gupta2012karma}
Shubham Gupta, Pedro Szekely, Craig~A Knoblock, Aman Goel, Mohsen Taheriyan,
  and Maria Muslea.
\newblock Karma: A system for mapping structured sources into the semantic web.
\newblock In {\em Extended Semantic Web Conference}, pages 430--434. Springer,
  2012.

\bibitem{housien2013comparison}
Hamed~Ibrahim Housien, Zhang Zuping, and Zainab~Qays Abdulhadi.
\newblock A comparison study of data scrubbing algorithms and frameworks in
  data warehousing.
\newblock {\em International Journal of Computer Applications}, 68(25), 2013.

\bibitem{kandel2011wrangler}
Sean Kandel, Andreas Paepcke, Joseph Hellerstein, and Jeffrey Heer.
\newblock Wrangler: Interactive visual specification of data transformation
  scripts.
\newblock In {\em Proceedings of the SIGCHI Conference on Human Factors in
  Computing Systems}, pages 3363--3372. ACM, 2011.

\bibitem{lionakis2017pfsgeo}
Panagiotis Lionakis and Yannis Tzitzikas.
\newblock Pfsgeo: Preference-enriched faceted search for geographical data.
\newblock In {\em OTM Confederated International Conferences" On the Move to
  Meaningful Internet Systems"}, pages 125--143. Springer, 2017.

\bibitem{maletic2005data}
Jonathan~I Maletic and Andrian Marcus.
\newblock Data cleansing.
\newblock In {\em Data Mining and Knowledge Discovery Handbook}, pages 21--36.
  Springer, 2005.

\bibitem{mountantonakis2018high}
Michalis Mountantonakis and Yannis Tzitzikas.
\newblock High performance methods for linked open data connectivity analytics.
\newblock {\em Information (2078-2489)}, 9(6), 2018.

\bibitem{papangelis2017ld}
Alexandros Papangelis, Panagiotis Papadakos, Margarita Kotti, Yannis Stylianou,
  Yannis Tzitzikas, and Dimitris Plexousakis.
\newblock Ld-sds: Towards an expressive spoken dialogue system based on
  linked-data.
\newblock In {\em Search Oriented Conversational AI, SCAI 17 Workshop
  (co-located with ICTIR 17)}, 2017.

\bibitem{raman2001potter}
Vijayshankar Raman and Joseph~M Hellerstein.
\newblock Potter's wheel: An interactive data cleaning system.
\newblock In {\em VLDB}, volume~1, pages 381--390, 2001.

\bibitem{sacco2009dynamic}
Giovanni~Maria Sacco and Yannis Tzitzikas.
\newblock {\em Dynamic taxonomies and faceted search: theory, practice, and
  experience}, volume~25.
\newblock Springer Science \& Business Media, 2009.

\bibitem{stonebraker2013data}
Michael Stonebraker, Daniel Bruckner, Ihab~F Ilyas, George Beskales, Mitch
  Cherniack, Stanley~B Zdonik, Alexander Pagan, and Shan Xu.
\newblock Data curation at scale: The data tamer system.
\newblock In {\em CIDR}, 2013.

\bibitem{tunkelang2009faceted}
Daniel Tunkelang.
\newblock Faceted search.
\newblock {\em Synthesis lectures on information concepts, retrieval, and
  services}, 1(1):1--80, 2009.

\bibitem{voting2016aid}
Y.~Tzitzikas and E.~Dimitrakis.
\newblock Preference-enriched faceted search for voting aid applications.
\newblock {\em IEEE Transactions on Emerging Topics in Computing}, PP(99):1--1,
  2016.

\bibitem{pfs2016species}
Yannis Tzitzikas, Nicolas Bailly, Panagiotis Papadakos, Nikos Minadakis, and
  George Nikitakis.
\newblock Using preference-enriched faceted search for species identification.
\newblock {\em International Journal of Metadata, Semantics and Ontologies},
  11(3):165--179, 2016.

\bibitem{tzitzikas2016faceted}
Yannis Tzitzikas, Nikos Manolis, and Panagiotis Papadakos.
\newblock {Faceted exploration of RDF/S datasets: a survey}.
\newblock {\em Journal of Intelligent Information Systems}, 2016.

\bibitem{CinderellaStickSpringer2018}
Yannis Tzitzikas and Yannis Marketakis.
\newblock {\em Cinderella's Stick - A Fairy Tale for Digital Preservation}.
\newblock Springer, 2018.

\bibitem{tzitzikas2012interactive}
Yannis Tzitzikas and Panagiotis Papadakos.
\newblock Interactive exploration of multidimensional and hierarchical
  information spaces with real-time preference elicitation.
\newblock {\em Fundamenta Informaticae}, 20:1--42, 2012.

\end{thebibliography}

\end{document}